\documentclass[aps,prl,reprint,groupedaddress,showpacs]{revtex4-1}
\usepackage{graphicx}
\usepackage{amsmath}
\usepackage{amsfonts}
\usepackage{amssymb}
\usepackage{wasysym}
\usepackage{color}

\begin{document}

\title{Fermi-Pasta-Ulam recurrence and modulation instability}
\author{E.A. Kuznetsov}
\affiliation{\textit{\small P.N. Lebedev Physical Institute, RAS, 53 Leninsky ave.,
                119991 Moscow, Russia;}\\
        \textit{\small L.D. Landau Institute for Theoretical Physics, RAS, 2 Kosygin
                str., 119334 Moscow, Russia;}\\
        \textit{\small Novosibirsk State University, 630090 Novosibirsk, Russia }}
                                
\begin{abstract}
We give a qualitative explanation of the analog of the Fermi-Pasta-Ulam
(FPU) recurrence in a one-dimensional focusing nonlinear
Schrodinger equation (NLSE). That  recurrence can be considered  as a result of the nonlinear development of 
modulation instability. All known  exact  localized solitons-type solutions describing 
 propagation on the background of the modulationally unstable
condensate show the recurrence to the condensate state after its interaction
with solitons. The condensate state locally recovers its original form
with the same amplitude but a different phase after soliton leave  its initial region.  This is the analog of the FPU recurrence for the NLSE.
Based on the integrability of the NLSE, we demonstrate that the FPU
recurrence takes place not only for condensate but for more general solution in the form of the cnoidal wave. This solution is periodic in space
and can be represented as a solitonic lattice. That lattice reduces to
isolated soliton solution in the limit of large distance between solitons.
The  lattice transforms
into the condensate in the opposite limit of dense soliton packing. The cnoidal wave is also modulationally unstable due to
soliton overlapping. This  instability at the linear stage does not
provide the cnoidal wave recurrence. The recurrence  happens at the nonlinear stage of the modulation instability.  From the
practical point of view the latter property is very important, especially for the
fiber communication systems which use soliton as an information carrier.
\end{abstract}

\maketitle

\vspace{0.5cm}

\textbf{1.} The phenomenon of recurrence in nonlinear systems with many
degrees of freedom was first observed in numerical experiment by Fermi,
Pasta and Ulam \cite{fermi1955studies} in 1954. The idea of Fermi  was to address how randomization due to the nonlinear
interaction leads to the energy equipartition between large number of
degrees of freedom in the mechanical chain. The chain in  \cite%
{fermi1955studies} had a quadratic nonlinearity and included $64$
oscillators supplemented with long-wave initial conditions. Instead of the
energy equipartition, numerical experiments showed that after a finite time
a recurrence to the initial data was achieved accompanied by a
quasi-periodic energy exchange between several initially exited modes. That
recurrence phenomenon became known as the Fermi-Pasta-Ulam (FPU) problem and
has been one of the most attractive subjects for numerous investigations.
Later, mainly by efforts of N. Zabusky,  these results were reproduced by means
of more powerful computers. Besides, there were observed many other
peculiarities in this problem (for details, see the original papers by
Zabusky (1962) \cite{zabusky1962exact}, Zabusky \& Kruskal (1965) \cite%
{zabusky1965interaction}, and Zabusky \& Deem (1967)\cite%
{zabusky1967dynamics} ). It was a time of forerunner of the era of
integrability for nonlinear systems.

Since the discovery of the Inverse Scattering Transform (IST), which was
first applied to the KDV equation by Gardner, Greene, Kruskal and Miura \cite%
{gardner1967method}, and later to the nonlinear Schrodinger equation (NLSE)
by Zakharov and Shabat \cite{shabat1972exact}, many aspects of the FPU
recurrence became more clear. In 1971 Zakharov and Faddeev \cite%
{zakharov1971korteweg} proved that the KDV equation, which, in particular,
can be obtained from the FPU system in the continuous limit for waves
propagated in one direction, represents completely integrable Hamiltonian
system. In 1974 Zakharov \cite{zakharov1974stochastization} demonstrated
that the so-called nonlinear string equation (sometimes called as the
Boussinesq equation) also belongs to the systems integrable by the IST. That
equation is close to  the continuous limit of the FPU system. According to 
\cite{zakharov1974stochastization} the long-time randomization for the FPU
system can be explained by the "distance" of that system to the nearest
fully integrable one. 
In such a case its dynamics will follow in accordance with the nearest
integrable system until deviations from the integrable
trajectory reach values of order 1. This time can be taken as an estimate
randomization time. 

Now one can find many review papers devoted to the FPU recurrence phenomenon (see,
e.g. \cite{ford1992fermi,zabusky2005fermi,porter2009fermi}). The main conjecture of all of
them is that the recurrence is a pure nonlinear phenomenon mainly intrinsic
to the integrable models. It is worth noting that the FPU recurrence was
also intensively studied experimentally. The first experimental
demonstration in optical fibers of the FPU recurrence was presented by Van
Simaeys \textit{et al.} \cite{van2001experimental} showing that the dynamics
of optical fields in the region of small third dispersion is described
with a good accuracy by the NLSE and connected with modulation instability
(MI). For current experimental situation, see \cite{mussot2014fermi} and
references therein.

The main aim of this paper is to explain how the recurrence phenomenon
appears within the one-dimensional NLSE  
\begin{equation}
i\psi _{t}+\psi _{xx}+2|\psi |^{2}\psi =0.  \label{NLS}
\end{equation}%
It is well-known that in many physical applications this equation can be
derived from the averaging of the equations of motion for solutions in the
form of quasi-monochromatic waves in a weakly nonlinear regime (see, e.g. \cite{zakharov1997hamiltonian}).
Therefore very often the NLSE is considered to be the envelope equation. In
the optical context, $\psi $ in Eq. (\ref{NLS}) is a dimensionless amplitude
of an electromagnetic wave packet in a reference frame moving with the group
velocity. It is well-known that this equation with a reasonable accuracy
describes propagation of optical solitons in fibers where they can be used
as an information carrier \cite{hasegawa1973transmission}. Note that the current
practical needs in fiber communications requires an increase in the
information rates and consequently a denser packing of information. In such
a case, soliton overlapping in soliton-based fiber communications becomes
very important factor. Fortunately, the NLSE has an exact solution in the
form of the soliton train, i.e. the so-called cnoidal wave. This is a whole
family depending on several parameters. In particular, the soliton solution
itself belongs to this family because it can be obtained from the cnoidal wave in
the limit of an infinite spatial period. Another limit of the cnoidal wave
is a solution in the form of a monochromatic wave or condensate. Such
solution is unstable with respect to the modulation instability, also known
as the Benjamin-Feir instability \cite{benjamin1967disintegration} (see,
also \cite{zakharov2009modulation}). Less is known about such instability
for the cnoidal wave \cite{kuznetsov1999modulation}. The growth rate in this
case can be found exactly by means of the linearized dressing procedure \cite%
{kuznetsov1984stability} and expressed in terms of $\sigma $- and $\zeta $-
Weierstrass functions. When the distance between solitons becomes large
enough, the maximal instability growth rate turns out to be exponentially
small, but increases with the distance decrease \cite%
{kuznetsov1999modulation}. For denser information packing the
modulation instability can destroy information and therefore one needs to
develop a nonlinear theory. From another side, a linear theory
can not explain the FSU recurrence in the NLS system. Here we explain the FSU recurrence from the fully nonlinear theory.

At the present time, there are known many exact solutions of the NLSE which describe
propagation of solitons/breathers on the condensate background.  Such
solution was constructed for a first time in \cite{kuznetsov1977solitons} and later in many
other papers (see \cite{peregrine1983water, akhmediev1985generation,
zakharov2013nonlinear} and references therein). All these solutions show
that after a while the condensate recovers its amplitude but has a different
(but constant) phase. This is the analog of the FPU recurrence for the NLSE.
In this paper we give a qualitative explanation of the FPU analog for
cnoidal waves. For fiber communications the latter means that such FPU
recurrence can ensure the preservation of information, in spite of the MI
existence.

\vspace{0.2cm}

\textbf{2.} We start from a stationary solution of Eq. (\ref{NLS}) in the
form {\ $\psi =e^{i\lambda ^{2}t}\psi _{0}(x)$} assuming for simplicity{\ $%
\psi _{0}(x)$} to be real. Then {$\psi _{0}(x)$} is defined from the Newton
equation, 
$
{\psi _{0}^{\prime \prime }}=-{\partial U}/{\partial \psi _{0}},
$
where $U=\frac{1}{2}\left( -\lambda ^{2}{\psi _{0}}^{2}+\psi _{0}^{4}\right) 
$ is the potential and $x$ has a meaning of time. In this case the
stationary NLSE has a first integral, i.e. the energy 
\begin{equation}
{\varepsilon =\frac{(\psi _{0}^{\prime })^{2}}{2}+U(\psi _{0})}
\label{energy}
\end{equation}%
that allows one to find a solution depending on two parameters ${\varepsilon 
}$ and $\lambda $. At $\varepsilon =0$ integration of this equation yields
the well-known soliton solution 
\begin{equation}
\psi _{0}=\lambda \,\mbox{sech}\lbrack \lambda (x-x^{(0)})\rbrack,  \label{soliton}
\end{equation}%
where ${x}^{(0)}$ is the coordinate of the soliton center of mass. (We remind that we consider here pure real $\psi _{0}$. In the general situation, soliton (\ref{soliton})  gets a constant phase multiplier.) 
When the
energy reaches minimal value in the potential ${U(\psi _{0})}$, ${%
\varepsilon }_{\min }=-\lambda ^{4}/8$, we arrive at another well-known
solution in the form of condensate with the constant amplitude,%
\begin{equation}
\label{condensate}
{\psi _{0}=\lambda /}\sqrt{2}.
\end{equation}  
Indeed, solutions \eqref{soliton} and \eqref{condensate} represent two limits of the two-parameters
family. This is the so-called cnoidal wave which can be expressed in terms
of elliptic functions. By introducing intensity $I=\psi _{0}^{2}$ and then
shifting, $I=-[\wp (x-i\omega ^{\prime })-{\lambda ^{2}/3}]$, Eq. (\ref%
{energy}) transforms into the equation for the elliptic Weierstrass
function: 
\begin{equation*}
(\wp ^{\prime })^{2}+U_{\wp }=0,  
\end{equation*}%
where 
$U_{\wp }=-4(\wp -e_{1})(\wp -e_{2})(\wp -e_{3})$
has a meaning of a new "potential energy" for trajectories related to a
(new) "energy" equal to zero. Here $e_{1,2,3}$ are values of $\wp $ in
points $z=\omega ,\omega +i\omega ^{\prime },i\omega ^{\prime }$, besides, $%
e_{1}>e_{2}>e_{3}$ and $e_{1}+e_{2}+e_{3}=0$. In the given case they are
equal to 
$
{e_{1}}=\lambda ^{2}/{3},\,{e_{2}}{=}-\lambda ^{2}/{6}+%
\sqrt{{\lambda ^{4}}/{4}+2{\varepsilon }},\,{e_{3}}{=}{-}{%
\lambda ^{2}}/{6}-\sqrt{{\lambda ^{4}}/{4}+2{\varepsilon }}.
$
The Weierstrass elliptic function is known as a double-periodic analytical function
with periods $2\omega $ (along real axis) and $2i\omega ^{\prime }$ (along
imaginary axis). Oscillations between zero points $e_{2}$ and $e_{3}$ in the
"potential" $U_{\wp }$ defines the real period $2\omega $. The oscillations between $e_{1}$ and $e_{2}$ in imaginary "time" ($%
x\rightarrow iy$) yields another period $2i\omega ^{\prime }$. As known \cite%
{whittaker1996course} (see also \cite{kuznetsov1974stability}), the
Weierstrass elliptic function can be represented in the form of the
solitonic lattice: 
\begin{eqnarray*}
&&\wp (x-i\omega ^{\prime })=-\mu ^{2}\Big\{ \sum_{n=-\infty }^{\infty }\mbox{sech}
^{2}\left[ \mu \left( x-2n\omega \right) \right] \label{Pe-1}\\ 
&& +2\sum_{n=1}^{\infty
}\mbox{cosech} ^{2}\left[ \mu 2n\omega \right] -\frac{1}{3} \Big\}, \nonumber
\end{eqnarray*}%
where $\mu =\pi /\left( 2\omega ^{\prime }\right) $. Respectively, for
the intensity $I$ we have 
\[
I=\mu ^{2}\sum_{n=-\infty }^{\infty }\left\{ \mbox{sech} ^{2}\left[ \mu \left(
x-2n\omega \right) \right] +\mbox{cosech} ^{2}\left[ \mu \left( 2n-1\right) \omega %
\right] \right\} . 
\]
In this lattice, the amplitude and inverse width for each soliton coincide and both
equal to $\mu $, in correspondence with (\ref{soliton}). Hence, one can
easily see that the soliton solution can be obtained as the limit of large
spatial period, $\omega /\omega ^{\prime }\rightarrow \infty $ when ${%
e_{1}\rightarrow e_{2}}$. 

Intensity $I$
reaches its minimum at the points $x_{n}=(2n+1)\omega $ corresponding to
half-distance between neighboring solitons: 
\begin{eqnarray*}
&&I_{\min }=\mu ^{2}\sum_{n=-\infty }^{\infty }\Big\{ \mbox{sech} ^{2}\left[ \mu
\left( x-2n\omega \right) \right] \\ &&+\mbox{cosech} ^{2}\left[ \mu \left( 2n-1\right)
\omega \right] \Big\} .
\end{eqnarray*}%
This constant pedestal is a result of overlapping between solitons. At large
distance between solitons, their overlapping is weak and, by this reason, $%
I_{\min }$ becomes exponentially small: 
\begin{equation*}
I_{\min }=\frac{4\pi ^{2}}{(\omega ^{\prime })^{2}}\exp \left( -\frac{\pi
\omega }{\omega ^{\prime }}\right) .
\end{equation*}%
In the opposite limit, $\omega ^{\prime }/\omega \rightarrow \infty $, when
the size of soliton in the lattice tends to infinity, the overlapping between
solitons becomes the main factor defining the cnoidal wave form: the function $%
\wp (x-i\omega ^{\prime })$ in this case tends to the constant value equal
to $-\lambda ^{2}/6$, corresponding to the condensate solution (\ref%
{condensate}), plus small harmonic oscillations, 
\begin{equation}
\wp (x-i\omega ^{\prime })\simeq -\lambda ^{2}/6+\sqrt{{\varepsilon }-{%
\varepsilon }_{\min }}\cos k_{0}x  \label{oscillations}
\end{equation}%
with $k_{0}=\pi /\omega =\sqrt{2}\lambda $. In this limit $e_{2}\rightarrow
e_{3}$.

\vspace{0.2cm}

\textbf{3.} It is well-known (see, e.g. \cite{zakharov2009modulation}) that the
condensate solution is unstable relative to the modulation instability
with  growth rate 
\begin{equation} \label{MI}
\gamma =k\sqrt{2\lambda ^{2}-k^{2}},
\end{equation}
 which 
 vanishes at $%
k_{0}=\sqrt{2}\lambda $ corresponding to stationary oscillations (\ref%
{oscillations}). It is less known that the cnoidal wave is also unstable relative
to the MI \cite{kuznetsov1999modulation}. To find the growth rate in this
case one needs to solve the NLSE linearized on the background of the
cnoidal wave by setting 
\begin{equation*}
\psi (x,t)=\psi _{0}(x)e^{i\lambda ^{2}t}+\phi,
\end{equation*}%
where $\phi $ is a small perturbation. Linearization of the NLSE gives the
system of coupled PDEs of the \textit{second} order. As it was shown in \cite%
{kuznetsov1999modulation}, using the IST simplifies significantly solution
of this linear problem, in particular, for the linearized NLSE 
this problem reduces to solution of the \textit{first} order PDEs. Following 
\cite{kuznetsov1999modulation}  we  here  show  how this system arises from the
auxiliary linear differential equations.

The NLSE can
be represented as a compatibility condition for two linear equations \cite{shabat1972exact}, 
\begin{eqnarray}
\frac{\partial \Psi }{\partial x} &=&L\Psi =i(\lambda \sigma _{3}+\hat{\psi}%
)\Psi ,  \label{compatibility} \\
\frac{\partial \Psi }{\partial t} &=&A\Psi =i(2\lambda ^{2}\sigma
_{3}+2\lambda \hat{\psi}+\hat{Q})\Psi \label{compatibility1},
\end{eqnarray}%
where $\Psi $ is the two-component vector function, $\lambda $ is the spectral
parameter and 
\begin{equation*}
\sigma _{3}=\left( 
\begin{array}{ll}
1 & 0 \\ 
0 & -1%
\end{array}%
\right) ,\;\;\hat{\psi}=\left( 
\begin{array}{ll}
0 & \psi ^{\ast } \\ 
\psi & 0%
\end{array}%
\right) ,\;\;\hat{Q}=\left( 
\begin{array}{ll}
-|\psi |^{2} & -i\psi _{x}^{\ast } \\ 
i\psi _{x} & |\psi |^{2}%
\end{array}%
\right) .
\end{equation*}%
As soon as $\psi $ satisfies the NLSE, a compatible solution of this
overdetermined system exists for all $\lambda$. Consider now two linear
equations for the $2\times 2$ matrix function $\Phi $: 
\begin{equation*}
\frac{\partial \Phi }{\partial x}=L\Phi -\Phi L,\;\;\frac{\partial \Phi }{%
\partial t}=A\Phi -\Phi A,
\end{equation*}
which are also compatible due to (\ref{compatibility}, \ref{compatibility1}). Then perturbation $\phi $ is defined by means of the relation 
\begin{equation*}
\left( 
\begin{array}{ll}
0 & \phi ^{\ast } \\ 
\phi & 0%
\end{array}%
\right) =[\sigma _{3},\Phi ].
\end{equation*}%
That can also be verified by direct calculation as follows. First, the diagonal part
of the compatibility equations and the spectral parameter $\lambda $ must be
excluded from this system. Second, a simple algebra then yields the linearized NLS
equation. 

This scheme is nothing more than the linearized version of the Zakharov-Shabat
dressing procedure \cite{zakharov1974scheme}. This version, in fact, was
introduced for a first time in 1974 \cite{kuznetsov1974stability} and developed
later in \cite{kuznetsov1984stability}. 

To find $\Phi $ for the cnoidal wave, one needs first to exclude a time dependence from the
matrices $L$ and $A$  by means of simple transformation
(rotation) making these matrices to be periodic functions in {\ $x$: $%
L\rightarrow L_{0}(x)$ and $A\rightarrow A_{0}(x)$. Thus, a solution of
this system is necessary to be sought in the form 
$\Phi _{0}(t,x)=\Theta (x)e^{\gamma t}$,
where $\Theta (x)$ is of the Bloch form with real quasi momentum $p$, 
$\Theta (x)=\theta _{p}e^{ipx},\;\;\theta _{p}(x+2\omega )=\theta _{p}(x)$.
Solvability condition to this transformed system gives the dispersion
relation $\gamma =\gamma (p)$ which is defined from a solution of  pure algebraic
equation 
\begin{equation*}
\gamma \Theta =[A_{0},\Theta ].
\end{equation*}%
Omitting then all quite simple calculations (for more details, see \cite%
{kuznetsov1999modulation}) we present the final answer for the maximal
growth rate when the distance between solitons becomes large enough: 
\begin{equation*}
\gamma _{\max }=8\left( \frac{\pi }{\omega ^{\prime }}\right) ^{2}\exp
\left( -\frac{\pi \omega }{\omega ^{\prime }}\right) .
\end{equation*}%
This expression shows that $\Gamma $  is exponentially small in this case but
grows with the spatial period decrease.

In another limit of the condensate, we arrive at the classical expression for
the MI growth rate  given by  (\ref{MI}).

\vspace{0.2cm}

\textbf{4.} What happens at the nonlinear stage of the MI for the cnoidal
wave? We know that, according to Zakharov and Shabat \cite{shabat1972exact}
(see, also \cite{novikov1984theory}), the phase space of the NLSE represents
discrete number of degrees of freedom which are solitons (these are the most
nonlinear objects) and solutions corresponding to the continuous spectrum.
Moreover, we know that collisions between solitons are elastic and
pairwise. A scattering of two solitons  results only in changing  of two their parameters, namely  the center of soliton mass  and its
phase as follows 
\begin{equation*}
\Delta x_{1}^{(0)}=\frac{1}{2\eta _{1}}\log \left\vert \frac{\lambda
_{1}-\lambda _{2}^{\ast }}{\lambda _{1}-\lambda _{2}}\right\vert ^{2},\,\,
\Delta \phi _{1}^{(0)}=2\,\mbox{arg}\left( \frac{\lambda _{1}-\lambda
_{2}^{\ast }}{\lambda _{1}-\lambda _{2}}\right) ,
\end{equation*}%
where $\Delta x_{1}^{(0)}$ and $\Delta \phi _{1}^{(0)}$ are respectively
shifts in mass center and phase for the first soliton,  $\lambda _{1,2}$ are
eigenvalues corresponding to the first and second solitons, $\eta =\mbox{Im}%
\,\lambda >0$ and $\lambda ^{\ast }$ means complex conjugation of $\lambda $.
Analogous formula can be written for the second soliton.

The cnoidal wave is of the form of the solitonic lattice. Therefore any
soliton from the lattice after interaction with a soliton propagating along
the cnoidal wave will undergo the same shift for its center of mass and
phase. This means that after scattering of the propagating soliton with the
lattice, the cnoidal wave will restore its previous form (up to the definite
spatial and phase shifts). Evidently, the same statement will be valid for
condensate as the partial solution of the cnoidal wave. The interaction of
condensate with any soliton after its propagation will restore amplitude of
the condensate but its (constant) phase will be different from the initial
value.

Scattering of a soliton with the non-soliton part also retains the soliton
form unchanged except  shifts of both center of mass of the soliton and its
phase. Thus, the cnoidal wave subject to the modulation instability, at
the nonlinear stage of the modulation instability development, should recover its form together with some phase and spatial
shifts.

This is the qualitative explanation of the FSU recurrence for the cnoidal
wave and for the condensate, in particular. It is necessary to underline
that the same phenomenon takes place for the KDV cnoidal wave that was found
by Mikhailov and the author in 1974 \cite{kuznetsov1974stability}. Similar to 
the KDV case, an exact solution describing propagation of soliton on the
cnoidal wave background can be found analytically for the NLSE also. For
instance, it can be done by means of the dressing procedure which  is however beyond the scope of
this paper. We remark only that this (still unknown) solution  will represent a
defect (or point dislocation) of the cnoidal wave. These objects are
non-stationary characterizing by two different frequencies. Like for the KDV
cnoidal wave, these defects propagate along the wave with some mean
velocity. In the partial case of the condensate these solitons were well
studied analytically (see, e.g. \cite{zakharov2013nonlinear} and references
therein).

\vspace{0.2cm}

\textbf{5.} In conclusion, we have presented qualitative explanation
for the analog of the FPU recurrence for the cnoidal waves in the presence
of perturbations, which are not assumed to be small. In this meaning the
recurrence for the condensate during nonlinear development of modulation instability
represents the partial case of the recurrence for the more general solution in the form of the cnoidal wave.  The
conjecture made in this paper is based on the integrability of the NLSE \cite{shabat1972exact}, in
particular, due to the elastic and pairwise  collisions between solitons and
between solitons and non-soliton part.
It is necessary to underline that the same phenomenon takes place for the
KDV cnoidal wave that was found in 1974  \cite%
{kuznetsov1974stability}. The arguments presented here for the NLSE can be
directly applied to explain recurrence for the KDV cnoidal wave.

The author thanks A.V. Mikhailov for discussions and P.M. Lushnikov for useful remarks. This work was supported
by the RSF (Grant No. 14-22-00174).

\end{document}